\documentclass[journal,letterpaper,10pt]{IEEEtran}
\usepackage{rbme-template}
\begin{document}
\title{A Roadmap for MEG Foundation Models}
\author{Philipp Thölke\textsuperscript{1,2}, Hamza Abdelhedi\textsuperscript{1,2,3}, Yorguin Mantilla-Ramos\textsuperscript{1,2,3}, Fouad Lbakali\textsuperscript{1,2,4}, Oumayma Gharbi\textsuperscript{3,5}, Catherine Duclos\textsuperscript{6,7,8}, Annalisa Pascarella\textsuperscript{9}, Vanessa Hadid\textsuperscript{2,10}, Oiwi Parker Jones\textsuperscript{11,12}, and Karim Jerbi\textsuperscript{1,2,3,13}
\thanks{\textsuperscript{1} Cognitive and Computational Neuroscience Laboratory (CoCo Lab), Université de Montréal, Montréal, QC, Canada\par \textsuperscript{2} Department of Psychology, Université de Montréal, Montréal, QC, Canada\par \textsuperscript{3} Mila -- Quebec Artificial Intelligence Institute, Montréal, QC, Canada\par \textsuperscript{4} IMT Atlantique, Brest, F-29238, France\par \textsuperscript{5} IVADO, Université de Montréal, Montréal, QC, Canada\par \textsuperscript{6} Center for Advanced Research in Sleep Medicine, Hôpital du Sacré-Cœur de Montréal, Santé Québec Nord-de-l'Île-de-Montréal - Universitaire, Montréal, QC, Canada\par \textsuperscript{7} Department of Anesthesiology and Pain Medicine, Faculty of Medicine, Université de Montréal, Montréal, QC, Canada\par \textsuperscript{8} Department of Neuroscience, Faculty of Medicine, Université de Montréal, Montréal, QC, Canada\par \textsuperscript{9} Institute for Applied Mathematics ``Mauro Picone'', National Research Council (CNR), Rome, Italy\par \textsuperscript{10} McGill University Health Centre, Montréal, QC, Canada\par \textsuperscript{11} Oxford Centre for Integrative Neuroimaging (OxCIN), University of Oxford, Oxford, United Kingdom\par \textsuperscript{12} Department of Engineering Science, University of Oxford, Oxford, United Kingdom\par \textsuperscript{13} UNIQUE Center, Quebec Neuro-AI Research Center, Montréal, QC, Canada}}
% Add the submission date and corresponding author's contact details when known.
\maketitle
\begin{abstract}
Foundation models are beginning to reshape brain-signal analysis by moving the field beyond task-specific decoding pipelines toward reusable models pretrained on broad neural datasets. Magnetoencephalography (MEG) is a compelling but still underdeveloped target for this shift: it captures human cortical dynamics at millisecond resolution while offering stronger spatial interpretability than EEG, making it especially valuable for source-resolved studies of perception, language, cognition, and clinical brain function. Yet MEG foundation models remain at an early stage, with only a small number of MEG-specific and MEG-inclusive multi-modal models, modest pretraining corpora, and emerging but still limited benchmarks. This perspective lays down the basic concepts needed to understand MEG foundation models and provides a didactic overview of the field’s key design choices, including tokenization, sensor- versus source-space representations, sensor-geometry encoding, backbone architectures, self-supervised objectives, and pretraining data. We then offer a roadmap for future development, organized around native MEG pretraining, adaptation of EEG foundation models, transfer from generic time-series models, and multi-modal integration with EEG, fMRI, MRI, behaviour, and stimulus features. We highlight the need for coordinated infrastructure, including diverse and reusable MEG datasets, rigorous evaluation across subjects, sites, tasks, and clinical settings, and responsible data-sharing practices that address consent, privacy, access, and governance.
\end{abstract}
\begin{IEEEkeywords}
brain decoding, foundation models, human electrophysiology, magnetoencephalography (MEG), multi-modal neuroimaging, neural representation learning, self-supervised learning.
\end{IEEEkeywords}

\section{Introduction}
\IEEEPARstart{T}{he} rise of foundation models in AI, most visibly through large language models, has triggered a paradigm shift that is now spreading beyond language, vision and speech into neuroscience and brain imaging~\cite{bommasani_opportunities_2022}. Rather than training a separate model for each task, foundation models learn from broad datasets and are then adapted to many downstream uses. Brain foundation models have so far advanced fastest where data are abundant or standardized, particularly in electroencephalography (EEG) and functional magnetic resonance imaging (fMRI), with EEG models such as Neuro-GPT and REVE illustrating the rapid emergence of reusable electrophysiological representations~\cite{cui_neuro-gpt_2024,ouahidi_reve_2025}. Magnetoencephalography (MEG) remains less developed, but early efforts such as MEG-GPT suggest that the same paradigm is beginning to reach MEG~\cite{huang_meg-gpt_2026}.

MEG combines millisecond temporal resolution with a closer link to cortical sources than EEG, because magnetic fields are only weakly distorted by the skull and scalp~\cite{hamalainen_magnetoencephalography---theory_1993,baillet_magnetoencephalography_2017}. This makes it uniquely valuable for modeling fast, distributed brain dynamics in perception, action, language, cognition and clinical neuroscience.

This article offers a didactic roadmap for MEG foundation models. We situate MEG within the broader landscape of brain foundation models, survey the early MEG-specific and MEG-inclusive efforts, and identify the design choices that will shape the field: tokenization, sensor- versus source-space representations, sensor-geometry encoding, backbone architectures, self-supervised objectives, pretraining data and benchmarks.

\subsection{What are foundation models?}

The foundation-model paradigm is easiest to understand from the example of language AI. Rather than training a separate model from scratch for each task, such as conversation, summarization, or classification, a large language model is first pretrained on broad and diverse text corpora to learn general-purpose representations. It can be then adapted and finetuned for a wide range of downstream uses. This replaces the older one-model-per-task approach with a more general strategy: learn reusable representations at scale, then re-purpose them across tasks and settings.

A foundation model is therefore not simply a large neural network, but a model trained on broad and heterogeneous data, usually with self-supervised objectives, to support tasks beyond those explicitly specified during training~\cite{bommasani_opportunities_2022}. Instead of relying only on human-provided labels, the model learns from the structure of the data itself by predicting information that has been masked or withheld.

For neuroscience, this raises a compelling possibility: models pretrained on large collections of neural time series may learn reusable representations of brain dynamics that generalize across subjects, tasks and recording contexts. For MEG, the central question is whether self-supervised pretraining on raw or minimally processed magnetic brain signals can yield such transferable representations of fast cortical dynamics. Rather than being trained to recognize one predefined outcome, such as a particular task or diagnosis, these models learn directly from the signals themselves, capturing recurring temporal patterns, spatial relationships, oscillatory dynamics, and interactions across brain regions. This is the “GPT effect” that brain-signal modeling is beginning to pursue: moving beyond task-specific classifiers toward general-purpose models whose learned representations can be adapted to many different neuroscientific and clinical questions.

\subsection{Background on EEG and fMRI foundation models}

Brain foundation models are now being developed across several neural data types, including EEG, fMRI, intracranial recordings and neural spiking data. MEG foundation models will therefore not emerge in isolation. Their development is shaped by neighbouring fields that are already further along, especially EEG and fMRI, and by growing efforts to build multi-modal models that learn shared representations across brain-imaging modalities.

EEG is the closest electrophysiological neighbour of MEG. It records scalp electric potentials with millisecond temporal resolution, is inexpensive and widely available, and has accumulated large public datasets across clinical, cognitive and brain–computer-interface settings. These features have made EEG the most mature domain for electrophysiological foundation models, with a rapidly growing literature that includes BENDR, BIOT, LaBraM, Neuro-GPT, NeuroLM and REVE~\cite{kostas_bendr_2021,yang_biot_2023,jiang_labram_2026,cui_neuro-gpt_2024,jiang_neurolm_2024,ouahidi_reve_2025}. Recent work is also extending this paradigm beyond EEG alone. SleepFM, for example, learned transferable representations from large-scale polysomnography by jointly modeling EEG alongside cardiac, muscular and respiratory signals, illustrating how neural time series can be embedded within broader multi-modal physiological foundation models~\cite{thapa_sleepfm_2024}. Here, we treat EEG as an important analog pathway for MEG foundation models, not as a direct comparison frame.

fMRI provides a different precedent. It measures a hemodynamic proxy of neural activity, with strong spatial resolution, large public repositories, standardized preprocessing pipelines and a natural anatomical coordinate system. Models such as BrainLM show how broad self-supervised pretraining can be applied to brain-imaging data with well-defined spatial structure~\cite{ortega_caro_brainlm_2024}.

MEG is now beginning to emerge as a distinct target for foundation-model development. Like EEG, it provides millisecond temporal resolution and can be analyzed both at the sensor level and, through anatomical modeling, in cortical source space. Recent work has pursued two complementary directions: MEG-specific foundation models and multi-modal models that integrate MEG with other neural signals. Recent examples include BrainOmni, which jointly models EEG and MEG; Brain-OF, which incorporates fMRI, EEG and MEG; and alignment frameworks that learn shared representations across EEG, MEG and fMRI~\cite{xiao_brainomni_2025,guo_brain-_2026,ferrante_towards_2026}. These efforts raise a central strategic question: should MEG be developed through dedicated MEG-native foundation models, or primarily as one component of broader multi-modal brain foundation models? The answer will likely depend on the downstream goal.

\subsection{The Case for MEG Foundation Models}

MEG records the magnetic fields generated by postsynaptic cortical currents. Like EEG, it captures brain activity at millisecond resolution; unlike EEG, these magnetic fields are less affected by the skull and scalp, giving MEG a closer link to cortical sources~\cite{hamalainen_magnetoencephalography---theory_1993,baillet_magnetoencephalography_2017}. This combination of high temporal resolution and spatial sensitivity makes MEG particularly well suited to modeling fast, spatially organized brain dynamics.

The case for MEG foundation models is therefore not that MEG will out-scale EEG, but that it can support a different class of questions. MEG is especially valuable when the target is not only when neural activity unfolds, but also where it originates and how it propagates across the cortex. A broadly pretrained MEG model could therefore learn representations useful for source-resolved dynamics, naturalistic language and cognition, clinical brain-state inference, and adaptation across subjects, stimuli, tasks and recording sites. For example, a pretrained MEG model could learn representations that distinguish activity arising from auditory versus language networks, track the millisecond-scale propagation of a spoken word from sensory to higher-order cortical regions, or characterize how large-scale network dynamics reorganize across sleep, anesthesia, or neurological disease. Because these representations are learned from large and diverse datasets rather than a single experiment, they could then be adapted to new participants, stimuli, cognitive tasks, clinical populations, or recording sites. The opportunity is therefore to build models that capture reusable spatiotemporal structure in human cortical dynamics (not simply reusable patterns in sensor-level signals).

The field remains early. A small number of MEG-specific models have appeared, including MEG-GPT, GPT2MEG and clinically oriented MEG pretraining, alongside MEG-inclusive multi-modal models such as BrainOmni, Brain-OF and EEG–MEG–fMRI alignment frameworks. Yet MEG pretraining corpora remain modest, benchmarking is only beginning to emerge, and the core design choices for MEG foundation models are still unsettled. This makes the field timely: the scientific rationale is clear, but the infrastructure and modeling principles still need to be built.

\begin{figure*}[h]
\centering
\includegraphics[width=\textwidth]{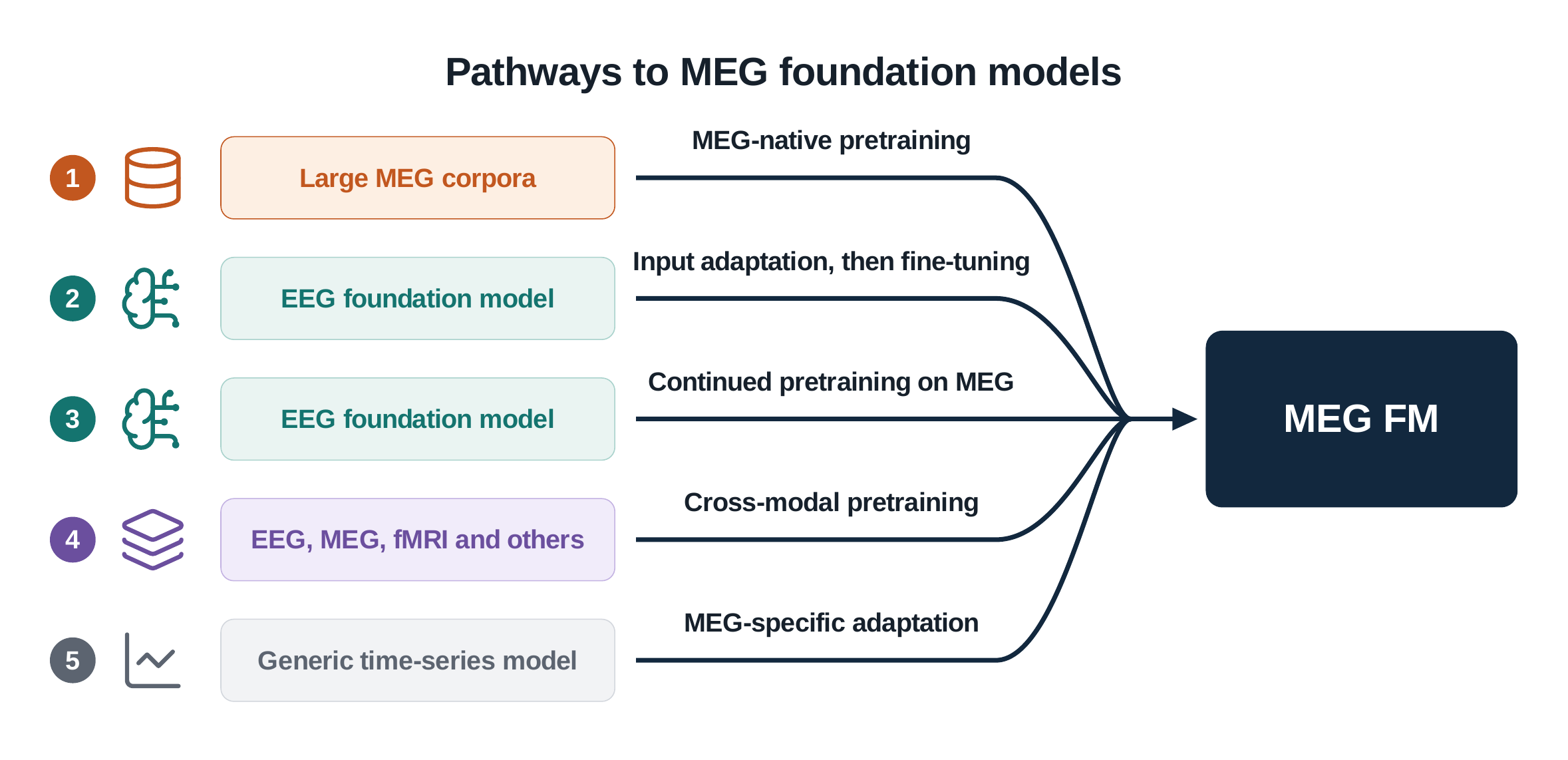}
\caption{\textbf{Pathways to MEG foundation models.} Complementary strategies for arriving at a reusable MEG foundation model (MEG FM), differing in where the model starts and what it must be taught about MEG. Each trades investment in data, compute and expressiveness against the others: routes that inherit from EEG or from generic time-series models demand less MEG data and less compute but constrain how much MEG-specific structure the model can express, whereas MEG-native pretraining is the most expressive and the most demanding. All aim at representations that generalize across subjects, tasks, recording sites and clinical settings.}\label{fig1}
\end{figure*}

\begin{sidewaystable*}
\caption{Current landscape of MEG foundation models and MEG-inclusive multi-modal foundation models. The table distinguishes direct MEG models, clinical MEG models, multi-modal MEG-inclusive foundation models, and precursors. For a list of public MEG datasets useful for pre-training and FM benchmarking, refer to our curated online resource at \url{https://github.com/thecocolab/meg-fm-resources}.}\label{tab1}
\footnotesize
\setlength{\tabcolsep}{3pt}
\begin{tabular}{@{}L{0.092\textheight}L{0.092\textheight}L{0.058\textheight}L{0.110\textheight}L{0.132\textheight}L{0.098\textheight}L{0.110\textheight}L{0.120\textheight}L{0.110\textheight}@{}}
\toprule
Model & Model type & Modalities & Data scale / datasets & Architecture & Pretraining objective & Tokenization / representation & Downstream evaluation & Key contribution \\
\midrule
MEG-GPT \cite{huang_meg-gpt_2026} & Direct MEG FM & MEG (resting-state) & Cam-CAN ($\sim$612 subjects) & GPT-style Transformer & Self-supervised autoregressive prediction & Region-level time-series tokenization & Cross-subject generalization, decoding & First explicit general-purpose MEG foundation model \\
\addlinespace
Headache-MEG-FM \cite{liao_pretrained_2026} & Clinical MEG FM & MEG (multi-state) & $\sim$416 participants (clinical + controls) & Deep neural network (details less standardized vs.\ GPT-style) & Self-supervised + supervised fine-tuning & Signal-level / feature embeddings & Migraine/headache classification & First clinically oriented MEG foundation model \\
\addlinespace
BrainOmni \cite{xiao_brainomni_2025} & Multi-modal FM & EEG + MEG & $\sim$1,997\,h EEG + 656\,h MEG & Transformer-based & Self-supervised cross-modal learning & Shared latent space across modalities & Cross-dataset generalization, decoding & First joint EEG--MEG foundation model \\
\addlinespace
Brain-OF \cite{guo_brain-_2026} & Multi-modal FM & fMRI + EEG + MEG & $\sim$40 datasets (multi-source) & Unified multi-modal architecture (transformer-like) & Self-supervised multi-modal pretraining & Cross-modal latent representations & Encoding, decoding, modality conversion & First tri-modal omnifunctional brain FM \\
\addlinespace
Ferrante et al.\ \cite{ferrante_towards_2026} & Hybrid FM (alignment-focused) & EEG + MEG + fMRI & Multiple datasets (vision tasks) & Representation alignment framework & Cross-modal alignment / shared embeddings & Latent representation learning & Decoding, encoding, modality translation & Early foundation-model-style alignment approach \\
\addlinespace
GPT2MEG \cite{csaky_gpt2meg_2026} & Precursor & MEG & Large MEG dataset (multi-subject) & GPT-2-style autoregressive model & Self-supervised autoregressive & Quantized MEG tokens & Generation, decoding & Key precursor to MEG-GPT paradigm \\
\addlinespace
Jayalath et al.\ \cite{jayalath_brains_2025} & SSL pretext-task method for MEG speech decoding (precursor; not a transformer FM) & MEG & Cam-CAN (641 subjects, $\sim$160\,h); + MOUS (204 subjects) in aggregation test & Convolutional cortex encoder (SEANet-style) + LSTM; dataset-conditional layer, FiLM subject conditioning; frozen backbone + linear probe & Self-supervised pretext prediction: band, phase-shift, amplitude-scale & Continuous multi-sensor windows (0.5\,s) conv-encoded to embeddings; sensor-count-agnostic & Speech detection + voicing classification on Armeni (3 subj.) and Gwilliams MEG-MASC (27 subj.); cross-subject, cross-dataset, novel-subject generalization & Neuroscience-inspired SSL tasks enabling first novel-subject and cross-dataset generalization in MEG speech decoding, with scaling-law evidence \\
\bottomrule
\end{tabular}
\end{sidewaystable*}

\section{Pathways to MEG foundation models}

There is no single route to a MEG foundation model. The field is still young, MEG data remain more limited than EEG or fMRI data, and different scientific goals will likely require different modeling strategies. This creates a choice between learning MEG-specific representations directly and leveraging the greater scale or complementary information available from other modalities and pretrained models. It is therefore useful to separate two strategic questions. The first is: where should the model come from? In other words, should it be pretrained directly on MEG, adapted from EEG, trained jointly across modalities, initialized from a generic time-series model, or embedded within a broader multi-modal architecture? Each route offers a different trade-off between MEG-specificity, scale, and transferability. The second is: in what space should MEG be represented? Should the model operate on sensor-space recordings, source-reconstructed signals, or both?

\subsection{Different Strategies for MEG foundation models}

\textbf{Strategy 1: MEG-native pretraining from scratch}

\noindent The most direct strategy is to train a foundation model on MEG data itself. In this route, both pretraining and downstream adaptation are designed around MEG-specific properties, including millisecond temporal resolution, sensor layout, source localization, vendor-specific acquisition systems and task structure. This is the cleanest route conceptually and the one most likely to capture MEG-specific structure. Early efforts such as MEG-GPT and GPT2MEG illustrate this direction. The limitation is practical: MEG-native pretraining requires broad, harmonized and reusable MEG corpora, which remain smaller and more fragmented than comparable EEG resources.

\medskip\noindent\textbf{Strategy 2: Transferring EEG foundation models to MEG}

\noindent The simplest transfer strategy is to take an existing EEG foundation model and adapt its input interface so that it can process MEG recordings, modifying channel metadata, input dimensions, sampling conventions or sensor mappings. Once the input is reconfigured, the adapted model can be used in three increasingly involved ways: extracting its embeddings directly, for instance through linear probing on a downstream task; freezing the backbone and training only a task-specific head on labeled MEG data; or fine-tuning the full model end to end. All three test whether representations learned from large-scale EEG serve as useful electrophysiological priors for MEG, which is particularly attractive when labeled MEG data are scarce. However, successful transfer is not guaranteed. EEG and MEG provide related but non-equivalent measurements of the same underlying neural currents: they differ in sensor geometry, spatial sensitivity and source orientation, as well as in signal scale, noise structure and preprocessing conventions. These differences create substantial domain shift, particularly for EEG models that encode modality-specific channel identities or spatial relationships. More fundamentally, EEG pretraining imposes an inductive bias: representations optimized to capture structure in EEG may not simply overlook MEG-specific information, but could constrain subsequent learning toward features that are suboptimal for MEG. This negative transfer could hinder the discovery of latent structure that is uniquely or more clearly expressed in MEG, particularly its richer spatial organization. EEG-to-MEG transfer should therefore be viewed as an empirical hypothesis rather than an inherently advantageous starting point, and benchmarked against MEG-native pretraining.

\medskip\noindent\textbf{Strategy 3: Continued pretraining of EEG foundation models on MEG data}

\noindent Continued pretraining goes one step beyond fine-tuning. In this route, an EEG foundation model undergoes additional self-supervised training on unlabeled MEG recordings before any supervised downstream task is introduced. Rather than being adapted immediately to a specific MEG task, the model first continues to learn from MEG itself, allowing representations initially learned from EEG to be refined for MEG.  This strategy may be one of the most practical near-term routes to MEG foundation models: it uses the scale advantage of EEG while allowing the model to adapt to MEG-specific temporal, spatial and sensor-level structure. Continued pretraining provides a natural bridge between transfer learning and MEG-native modeling while potentially requiring less MEG data to learn robust representations by leveraging latent structure already learned from EEG. The extent to which EEG-derived representations facilitate this adaptation, or remain influential after continued MEG pretraining, is an important empirical question that can be addressed through comparison with MEG-native models.

\medskip\noindent\textbf{Strategy 4: Cross-modal pretraining including MEG}

\noindent Rather than treating EEG as a source model and MEG as a target domain, MEG can be included in pretraining from the outset alongside other modalities. In its narrower form this means joint EEG--MEG pretraining, treating the two as related views of fast electrophysiological activity. Whereas continued pretraining likely results in the loss of EEG-specific representations, a jointly trained model can learn which features are shared while preserving modality-specific information where needed. BrainOmni is an example of this direction, using a unified framework for EEG and MEG and showing that joint pretraining outperforms single-modality models on both.

The broader form extends the same logic to modalities with different spatial, temporal and biophysical properties: fMRI for spatial organization, MRI for anatomy, and stimulus-related, behavioural or language variables for task context, as in Brain-OF and in alignment frameworks linking EEG, MEG and fMRI representations. This strategy is appealing because it exploits the larger scale of neighbouring modalities while keeping MEG in the pretraining objective, and it suits scientific questions that require complementary information. Its main challenge is asymmetry: ensuring that the learned representation is not dominated by the more abundant modality and that MEG contributes meaningfully rather than becoming a peripheral input.

\medskip\noindent\textbf{Strategy 5: Adapting generic time-series foundation models}

\noindent A final route is to adapt foundation models developed for generic time-series data, such as Chronos-2~\cite{ansari_chronos-2_2025} or MOMENT~\cite{goswami_moment_2024}. These models may provide useful architectures, long-context sequence handling or scalable pretraining recipes. However, MEG is not a generic time series. It has sensor geometry, channel-type structure, physiological artifacts, sampling conventions and neurobiological constraints that must be incorporated into the model or its input representation. Generic time-series models may therefore offer useful starting points, but they require MEG-specific adaptation before they can serve as credible MEG foundation models. A related strategy is to transfer foundation models pretrained on other non-neural time-series modalities whose structure may be relevant to a specific MEG task. For example, NeuSpeech and MAD leverage representations from the audio-pretrained Whisper model for MEG-to-speech or MEG-to-text decoding~\cite{yang_mad_2025,yang_neuspeech_2026}. Such approaches differ from generic time-series transfer in that they exploit modality-specific representations learned from an external signal domain rather than a domain-agnostic time-series model.

\subsection{Sensor- and source-space representations}

A second strategic decision concerns how MEG data are represented before entering the model. This choice is partly independent of the model’s starting point. A MEG-native model, an EEG-adapted model or a multi-modal model could, in principle, operate on sensor-space or source-space signals.

Sensor-space models operate directly on the recorded MEG channels. They remain close to the measured data, avoid committing to a particular inverse solution, and are well suited to large-scale pretraining on raw or minimally processed recordings. Their main difficulty is that MEG sensor layouts vary across systems, vendors and magnetometer configurations. A sensor-space foundation model must therefore handle differences in channel number, sensor position, sensor orientation, channel type and missing sensors. One promising solution is to augment model inputs with channel geometry via positional encoding, allowing the model to learn spatial features. Furthermore, Défossez et al.~\cite{defossez_decoding_2023} introduced a geometry-aware spatial attention mechanism intended to support recordings with different sensor layouts, although its specific contribution to cross-dataset performance was not isolated experimentally. Positional encoding of sensor geometry has also found success in recent EEG foundation models~\cite{ouahidi_reve_2025,jiang_large_2023,yang_biot_2023,wang_cbramod_2025}.

Source-space models first project MEG data onto cortical locations, regions or parcels. This can make the learned representations more anatomically interpretable and may help align recordings acquired with different MEG systems into a common brain space. Source-space modeling is attractive for cognitive neuroscience, where the goal is often to relate model representations to cortical regions, networks or source-level dynamics. The tradeoff is that source reconstruction introduces assumptions about the  head model, noise covariance, regularization and requires an anatomical MRI scan for every subject. A further consideration is that source reconstruction can itself impose common spatial structure across datasets through the inverse method, regularization and parcellation scheme. Apparent cross-system alignment in source space may therefore reflect a mixture of shared cortical organization and shared reconstruction assumptions, motivating robustness tests across inverse solutions and spatial representations.

\subsection{Choosing the best strategy}

No single strategy will dominate all use cases for MEG foundation models. The best route will depend on the scientific question, the available data, the desired level of anatomical interpretability, and the evaluation setting. A clinical classifier may benefit from continued pretraining followed by supervised fine-tuning on disease-specific data. A naturalistic language or speech model may require MEG-native pretraining, multi-modal brain–stimulus alignment, or continued pretraining on richly annotated auditory datasets. A cognitive-neuroscience model aimed at interpreting cortical dynamics may benefit from source-space or hybrid representations. Selecting a strategy involves trade-offs between compute requirements, implementation complexity, and expected performance. While cross-modal pretraining (Strategy 4) offers a desirable pairing of modalities, thus enabling more advanced analyses, this approach implies complex pipelines to process multi-modal data. Conversely, fine-tuning or continued pre-training of existing EEG models (Strategies 2 and 3) risks yielding lower performance, but offers a significantly lighter computational footprint and lowers the barrier to implementation. The central question is therefore not which pathway is universally best, but which combination of initialization, representation space and training objective best serves the target scientific problem.

\section{Key ingredients of a MEG foundation model}

We organize this section around the native-MEG strategy, in which pretraining and downstream evaluation both use MEG, and close with pointers on adapting foundation-model approaches from neighbouring modalities such as EEG. The recurring theme is representation. How MEG is presented to the model must suit both the backbone architecture and the pretraining objective, and choices along this axis propagate to every downstream property: cross-system transferability, sample efficiency, the structure the model can recover, and the tasks it can solve.

\subsection{Data input and tokenization}

The first question is how to segment raw MEG, an array of channels or a representation derived from voxels (in the case of source-reconstructed data) over time, into units (tokens) the model can ingest. This has a spatial and a temporal component.

Spatially, sensor channels are the natural unit in sensor space. In source space, an ROI strategy is needed to keep the context size tractable, either by aggregating over a predefined anatomical atlas or by learning the aggregation from data. MEG-GPT parcellates source activity into 52 atlas-based ROIs, whereas most other models operate directly in sensor space. Data-driven cortical parcellation has previously been explored for source-reconstructed EEG/MEG, for example using adaptive parcellations derived from functional connectivity structure~\cite{farahibozorg_adaptive_2018}, although how best to learn spatial tokenization specifically for MEG foundation models remains largely open. 

Temporally, the dominant strategy for transformer backbones is to patch the signal into tokens spanning a few hundred milliseconds to several seconds, with no clear consensus yet on the optimal temporal scale. Architectures that are not transformers usually avoid fixed patching and instead consume continuous data through convolutional or recurrent encoders. MEG-GPT uses a hybrid tokenizer that maps raw signals to discrete codebook tokens via convolution and recurrence, BrainOmni uses a convolution-based tokenizer, and BBL (Jayalath et al.) is convolutional with a final recurrent layer. Recent systematic comparisons of sample-level MEG tokenization further suggest that simple fixed discretization schemes can perform comparably to learned tokenizers across several evaluation settings, indicating that greater tokenization complexity may not always be necessary~\cite{cho_systematic_2026}.

One clear advantage of MEG over EEG is the ability to better map sensor-level recordings back to anatomical sources. A foundation model would ideally be able to encode this greater spatial precision, making source-space tokenization not merely a way to bound context size but a route to representations tied to cortical anatomy.

\subsection{Backbone architectures}

The backbone is the main representation-learning component of the model: it transforms  input tokens into contextual representations that can be used for latent analysis and downstream tasks, and typically contains the majority of the model’s trainable parameters. Transformers have become the dominant backbone  because they scale effectively and can capture relationships across long sequences. They also dominate the emerging MEG foundation-model landscape, though they appear with several adaptations, tailored to the particular challenges  of neural data (e.g. long recordings, many sensors, heterogeneous modalities). One important adaptation is the use of perceiver-style modules, which compress large or heterogeneous inputs into a smaller, fixed set of latent representations before further processing. This reduces computational demands and can provide a common representation for inputs that differ in their number of sensors, regions or modalities. MEG-GPT, for example, combines a decoder-only transformer with an autoregressive perceiver to reduce the computational and memory costs of autoregressive modeling. BrainOmni uses a Criss-Cross Transformer to model spatial and temporal dependencies separately. Brain-OF combines DINT attention with a sparse mixture-of-experts backbone, and its perceiver-style ARNESS module first maps heterogeneous inputs into a common set of latent tokens to facilitate multimodal integration. BBL provides the main non-transformer contrast, using a SEANet-inspired convolutional backbone.

The prevalence of transformers is consistent with their success across language, vision, and other sequential domains, but it should not be taken to imply that they are inherently optimal for MEG. Non-transformer-based convolutional architectures remain highly effective for continuous signals such as audio, leaving substantial room for alternative backbones tailored to the temporal and spatial structure of raw MEG.

\subsection{Positional encoding and sensor awareness}

By themselves, transformers are permutation-equivariant, which means that the order of tokens is preserved throughout the model but does not have any influence on the computation happening inside of the model. To inform the model of both the temporal and spatial ordering of tokens, positions must be encoded explicitly. Spatial encoding creates a challenge for MEG as representations differ between sensor and source space, and vendors differ in both channel layout and channel type (magnetometers and gradiometers). Positional encoding is primarily a transformer-specific concern; convolutional encoders such as BBL do not require it, because they operate on continuous data and are typically tied to a fixed channel configuration.

In current MEG foundation models, temporal position encoding is commonly handled through learned embeddings (MEG-GPT, GPT2MEG), absolute sinusoidal encodings (Headache-MEG-FM), or RoPE (BrainOmni, Brain-OF). For spatial positions, the existing schemes follow two directions: categorical sensor/parcel identity for single-system models (GPT2MEG, MEG-GPT), and full 3D position together with orientation and sensor-type embeddings for the cross-vendor setting (BrainOmni). The most general of these schemes is typically preferable, since the goal is one model that transfers across recording systems rather than one restricted to a single vendor. This issue may become even more important with optically pumped magnetometer (OPM) systems, where wearable and reconfigurable sensor arrays can introduce greater variability in sensor-to-head geometry and movement. Future MEG foundation models may therefore need spatial representations that accommodate dynamic geometry rather than treating sensor position as a fixed property of the recording system.

\subsection{Self-supervised objectives}

Supervised machine learning requires labeled data samples, but the emergence of self-supervised learning has unlocked the use of vast amounts of unlabeled data. MEG data does in fact carry labels in many cases, since recordings are often task- or condition-specific, yet self-supervised pretraining still helps to ingest data in a unified manner and improves generalization. The objective in every case is the same: to create a general model of the given domain without relying on annotations, the model predicts withheld or transformed views of its own input, or in the case of cross-modal alignment, matches its representations to a paired complementary modality such as images, audio or text.

Self-supervised strategies differ but fall into a few general classes: masked signal prediction in raw signal or latent space (e.g. JEPA); autoregression; contrastive learning (e.g. SimCLR); cross-modal alignment, which is a special case of contrastive learning (e.g. CLIP); and prediction of surrogate features. Current MEG-FMs populate several of these. Masked reconstruction appears in Headache-MEG-FM (a bidirectional masked autoencoder with MSE loss), BrainOmni (masked codebook prediction), and Brain-OF (joint masked prediction in time and frequency). Autoregressive next-token prediction is used by MEG-GPT (over ROI codes) and GPT2MEG. Cross-modal alignment is used by Ferrante et al. (CLIP-style InfoNCE against frozen CLIP image embeddings). Surrogate-feature prediction is used by BBL (spectral power, phase-shift, and amplitude). Notably, several classes still have no MEG instance: latent-space masked reconstruction (JEPA), spectral reconstruction, and contrastive predictive coding.

\subsection{Pretraining data}

The pretraining corpus is a central determinant of what a foundation model can learn and how well its representations generalize. Current MEG foundation models are typically pretrained on a single dataset or experimental paradigm (MEG-GPT, GPT2MEG, Headache-MEG-FM), which limits the breadth of the representations they can learn. This contrasts with foundation models in other domains, which often rely on large and heterogeneous pretraining corpora.
For MEG, building such resources requires attention not only to data scale, but also to data quality and composition. Work in language modeling has shown that carefully curated, high-quality data can substantially improve learning efficiency and performance, challenging the assumption that increasing dataset size alone is sufficient~\cite{gunasekar_textbooks_2023,chen_revisiting_2025}. A useful MEG pretraining corpus should therefore span the populations, tasks, recording systems, and experimental conditions relevant to its intended downstream applications, while avoiding excessive representation of any single setting. These factors may strongly influence cross-dataset and cross-condition transfer, yet their contribution to MEG foundation-model generalization remains largely unexplored. Early evidence already suggests that corpus composition matters: selecting datasets for aggregation based on their standalone decodability can outperform indiscriminately pooling all available data~\cite{jayalath_brains_2025}. Developing larger, more heterogeneous MEG corpora, together with principled strategies for their curation, selection and balancing, will therefore be an important step toward broadly generalizable MEG foundation models. A compiled list of currently available MEG datasets with information on number of participants, public availability, modalities and data scale can be found at \url{https://github.com/thecocolab/meg-fm-resources}.

\begin{table*}[!t]
\caption{Foundation-model terminology}\label{glossary}
\centering
\begin{tabular}{@{}p{0.22\textwidth}p{0.75\textwidth}@{}}
\toprule
Term & Definition \\
\midrule
\multicolumn{2}{@{}l}{\textit{Core concepts for MEG foundation models}} \\
\midrule
Foundation model & A general-purpose model pretrained on broad and diverse data, then adapted to many downstream tasks rather than trained from scratch for each new task. \\
Pretraining & The initial large-scale training stage, usually on mostly unlabeled data, in which the model learns general representations before being adapted to specific tasks. \\
Fine-tuning & Updating some or all of a pretrained model's parameters on a downstream task so that the model becomes specialized for a new dataset, question, or application. \\
Linear probing & Evaluating a pretrained model by freezing its internal representations and training only a simple linear classifier or regressor on top of them. \\
Downstream task & Any task performed after pretraining, such as cognitive-state decoding, clinical classification, speech decoding, or behavioral prediction. \\
Transfer learning & Reusing knowledge learned in one setting, such as another dataset, task, or modality, to improve performance in a different setting, including EEG-to-MEG adaptation. \\
Embedding / representation & A learned vector that summarizes the signal content of a token, patch, or trial in a form that can be reused for prediction, comparison, or transfer. \\
Tokenization & The process of segmenting continuous MEG data into units the model can ingest, such as temporal patches, source-region time series, or discrete codes. \\
Sensor-space vs.\ source-space representation & Two ways of representing MEG data: sensor space refers to signals at the recorded sensors, whereas source space refers to signals projected onto cortical locations or regions. \\

\midrule
\multicolumn{2}{@{}l}{\textit{Self-supervised training strategies for MEG foundation models}} \\
\midrule
Self-supervised learning & A training paradigm in which models learn from unlabeled data by solving prediction problems derived from the data itself, rather than relying on human-provided labels. \\
Masked modeling & A self-supervised strategy in which part of the input is hidden and the model is trained to predict the missing content from the surrounding context. \\
Autoregressive forecasting & A training strategy in which the model predicts the next token or segment from the past, encouraging it to learn temporal structure and sequence dependencies. \\
Contrastive learning & A representation-learning strategy that brings related examples closer together in embedding space while pushing unrelated examples farther apart. \\
JEPA / latent predictive learning & A predictive approach in which the model learns to predict the latent representation of missing or future content rather than reconstructing the raw signal itself. \\
Reconstruction space & The space in which the training loss is computed, for example raw signal space, spectral space, latent space, or discrete code space. \\

\bottomrule
\end{tabular}
\end{table*}

\section{Assessing the performance of MEG FMs}

Whether MEG foundation models succeed is ultimately an empirical question: does large-scale pretraining yield representations that transfer across subjects, tasks and recording sites, and by what measure would we know? And, how can black-box foundation models trained on neuroimaging data yield neuroscientific insights beyond classical methods? For now, answers to these questions are unclear for two reasons. First, the parallel EEG literature counsels caution about the scaling argument often made on their behalf: across matched protocols, larger EEG models have not reliably generalized better than smaller ones~\cite{liu_eeg-fm-compass_2026,kuruppu_eeg_2026}, tempering the expectation that scale alone will carry MEG. Second, MEG-FMs have until recently been evaluated on idiosyncratic splits of single datasets, each paper choosing its own task, cohort and metric, leaving results incomparable across models. Standardized, leaderboarded benchmarks have only begun to appear, and the evidence base they provide remains thin.

\subsection{Existing benchmarks and evaluations for MEG FMs}

The first leaderboarded MEG benchmark comes from the 2025 PNPL competition~\cite{landau_2025_2025}, which is built on LibriBrain~\cite{ozdogan_libribrain_2025}, a deep single-participant corpus of more than 50 hours of MEG recorded while one subject listened to English audiobooks and providing standardized data splits. The competition defines two tasks, speech detection and phoneme classification. The 2026 competition~\cite{mantegna_libribrain100_2026} progresses to word classification and between-subject evaluation, building on the LibriBrain100 dataset: $\sim$104 hours of MEG acquired across 33 subjects ($\sim$80 hours from one subject and $\sim$40 minutes from 32 others). This design makes use of the insight that decoding performance improves faster when data is scaled up within- rather than between-subjects (see e.g.~d’Ascoli et al.~\cite{dascoli_towards_2025}). Its baseline model MEG-XL~\cite{jayalath_meg-xl_2026} demonstrates that a model trained on large within-subject data can generalize to new subjects with only a small amount of fine-tuning data. From a foundation modeling perspective, the main limitation of these benchmarks is their relatively narrow focus on speech decoding. However, they do provide a useful set of resources to evaluate foundation models, provided that similar benchmarks can be established for tasks across a wider range of domains. 

A second MEG-relevant effort is NeuralBench~\cite{banville_neuralbench_2026}, a unifying open-source framework designed to benchmark general NeuroAI models. NeuralBench sits within a broader wave of foundation-model benchmarking efforts in electrophysiology, including the EEG-focused AdaBrain-Bench, EEG-FM-Bench, OpenEEG-Bench and NeuroAtlas~\cite{wu_adabrain-bench_2025,xiong_eeg-fm-bench_2026,guetschel_toward_2026,kontras_neuroatlas_2026}. These efforts differ in scope, emphasizing, for example, BCI applications, standardized evaluation across EEG paradigms, community-driven benchmarking, or clinical generalization, but currently focus primarily on EEG.

NeuralBench is broader in modality and, unlike the PNPL competitions, was designed with general foundation models in mind. Its first release is EEG-primary, spanning 36 tasks across 94 datasets, while MEG is included as a proof-of-concept extension with two tasks: image decoding on the public THINGS-MEG dataset and keystroke decoding using the Lévy et al.~\cite{levy_noninvasive_2026} dataset, which is available to academic researchers upon request. Its most suggestive MEG result is that REVE, pretrained on EEG alone, performs best among the tested models on the MEG typing task, providing early evidence that electrophysiological representations may transfer across modalities. As a specific MEG benchmark, however, NeuralBench remains preliminary, with only two tasks and limited task coverage.

While benchmarking may still benefit from wider task coverage (clinical, subject metadata, cognitive tasks), the move toward standardized data splits and public leaderboards encourages better comparability across methods. PNPL drives progress in speech decoding, providing increasingly large benchmarking data with a maturing tutorial and leaderboard infrastructure and may prove to play an important role in MEG foundation model evaluation.

\subsection{What standardized MEG-FM benchmarking would need}

A benchmark is diagnostic only to the extent that its structure mirrors the claim being tested. The foundation-model claim is that self-supervised pretraining yields broad, reusable representations that transfer across conditions a single-task model cannot. A benchmark that holds those conditions fixed cannot test the claim, however large its leaderboard.

Concerning data, benchmarks must be built on open datasets so that results are reproducible and models comparable across groups; a suite resting partly on private data, as NeuralBench's MEG component currently does, cannot serve as a shared standard. It must span the diversity the model claims to generalize over: cognitive and clinical conditions, resting and task states, and different acquisition systems and vendors. Differences in sensor geometry and channel type currently prevent usage of some MEG FMs on datasets acquired using a different MEG system, though this is not an inherent barrier: geometry-aware encodings (BrainOmni) and pretraining across multiple acquisition systems~\cite{jayalath_brains_2025,jayalath_meg-xl_2026} have both been shown to support transfer across different MEG systems. Generalization should also be distinguished from longitudinal reliability. For applications that depend on individual-level representations, particularly clinical or longitudinal ones, models should be evaluated on whether relevant features remain stable across repeated recordings of the same individual despite session-to-session variation in head position, coregistration and measurement conditions. Finally, benchmark data must be excluded from pre- and post-training stages and used purely for downstream evaluation. In increasingly aggregated neuroimaging corpora, this exclusion should ideally be verified at the participant and recording-session level, since the same recordings or derived versions of them may appear across multiple releases or datasets. This is standard practice but due to the scarcity of MEG data, pretraining benefits from using as much data as possible, inherently limiting the feasibility of large-scale evaluation datasets.

Comparability also requires a standardized evaluation methodology, since approaches that can be applied to every model should not be mixed across them. Linear probing measures how linearly separable the conditions are in embedding space, while full fine-tuning permits non-linear transformation of the embeddings before classification or regression and so tends to score higher. Reporting only fine-tuned numbers can therefore overestimate scores of a model whose frozen features are not linearly separable. Foundation models must further be compared against matched baselines under the same protocol, both hand-crafted feature pipelines and task-specific deep networks. This bar matters because MEG specialists are well developed, from EEGNet ports to Meta's brain-decoding architectures, and because the one broad comparison available so far, from NeuralBench on EEG, finds foundation models only marginally ahead of task-specific ones. Building a suite that meets these five requirements is not incidental to the MEG foundation-model program but the precondition for its central claim to be testable at all.

Finally, progress will require evaluating MEG foundation models not only as predictors of downstream conditions but also as general models of brain activity. Their scientific value can be assessed by whether patterns identified in their learned representations generate new hypotheses, replicate across independent datasets, and withstand experimental validation. Applying these models to concrete neuroscientific questions will be essential for developing rigorous methods that translate their internal representations into reproducible scientific insight.

\section{Open questions and future paths for MEG FMs}

The previous sections map where MEG foundation models currently stand. What follows is a roadmap: the concrete problems whose resolution would move the field from early demonstrations to reusable tools, ordered from data and modeling to the deeper question of what these models are ultimately for.

\subsection{Building usable, diverse pretraining corpora}

The pretraining corpus is among the main constraints on MEG foundation models. Composition may matter as much as raw volume, but MEG is currently limited on both fronts. A plausible near-term goal is an open, aggregated MEG corpus spanning thousands of subjects and hours, drawn from many sites and vendors, offering a common ground for the development of MEG foundation models.

Such aggregation efforts ask for a certain level of standardization across data organization and representation. Preprocessing should be kept minimal, following the light-touch convention emerging in EEG foundation models, since heavy or idiosyncratic pipelines both discard signal the model might learn from and tie the corpus to choices that may not transfer across sites. A distinct MEG decision is whether to source-reconstruct: projecting to cortical space can improve cross-system comparability and anatomical interpretability, at the cost of the forward- and head-model assumptions that source reconstruction introduces.

Finally, the corpus must be built with evaluation in mind. Any dataset reserved for benchmarking must be excluded from pretraining, so corpus construction and benchmark construction have to be coordinated from the outset rather than reconciled after the fact. Given MEG's data scarcity, a majority of datasets likely end up being part of the pretraining corpus, limiting MEG foundation model evaluation to only low data regimes.

\subsection{Pretraining objectives matched to MEG's structure}

The pretraining objective determines what structure the model learns, and to which it stays invariant. More generally, useful invariance does not imply removing all variability across individuals or recordings. For MEG foundation models, the challenge is therefore to suppress variability related to measurement and acquisition while preserving meaningful inter-individual differences that may carry neuroscientific or clinical information. Table \ref{glossary} provides an overview of commonly used self-supervision objectives: masked modeling, with the reconstruction loss computed against raw signal, spectral, latent or codebook predictions; autoregressive next-token prediction; and alignment against a paired modality such as stimulus features or behaviour. Current MEG models populate only part of this space, leaving latent-space prediction, spectral reconstruction and contrastive predictive coding underexplored.

Which of these suits MEG is partly a question of where the predictable structure lies. Single-trial MEG has low signal-to-noise, so a loss computed in raw signal space spends much of the model's capacity on content that is not predictable in principle (noise). Both existing MEG models that predict autoregressively (MEG-GPT and GPT2MEG) avoid this by quantizing the signal first. Alignment objectives occupy a different position: they require paired streams with contrastive representations of the same underlying dynamics.

The most likely near-term source of evidence is EEG, where a wider range of objectives has been tried at a larger scale. Masked reconstruction dominates there, with models predicting held-out content in signal, spectral, latent or codebook space. The reconstruction space itself has not settled on a single method yet. Latent-space prediction has reported competitive cross-subject performance at substantially smaller scales than approaches reconstructing raw data. These are a reasonable prior rather than a prediction, particularly through the option to define the objective in source space. In both modalities the decisive comparison at matched data, compute and architecture is still missing, without which the choice of objective remains unclear.

\subsection{Tokenization and continuous-time representations}

While MEG and EEG data are both multivariate oscillatory time-series, there are some relevant differences for tokenization and positional encoding. A common strategy for tokenization in EEG foundation models is splitting the raw signal into short patches, typically around one second, and positionally encoding each channel by electrode location on the scalp. In sensor space MEG can mostly follow this approach, since the underlying problem of variable montages across sites and devices is shared. Depending on the vendor, however, a single sensor position may carry a magnetometer and two orthogonal planar gradiometers, measuring different quantities in different units at the same point in space. Position alone is therefore insufficient to identify a channel, and sensor orientation and channel type have to enter the positional encoding alongside it.

It is still an open question whether raw patch embeddings are the best tokenization approach for either EEG or MEG and some EEG efforts tokenize spectral rather than time-domain content, or follow convolutional approaches rather than patches. Adaptive or data-driven tokenization, in which patch boundaries are predicted by the model rather than fixed in advance, remains essentially unexplored in EEG and untried in MEG. Recent efforts by Pradeepkumar et al.~\cite{pradeepkumar_tokenizing_2026} have explored learning a vector-quantization-based discrete vocabulary of tokens for single-channel EEG, suggesting potential gains arising from learned time-frequency tokens over patch embeddings.

Source space changes the scale of the problem. Source reconstruction expands a few hundred sensors into thousands or tens of thousands of cortical locations, depending on the resolution of the source grid. Since context size of the models is bounded, this forces a spatial tokenization choice. Atlas-based parcellation is the simple option, aggregating source activity into a predefined set of anatomical regions, which fixes spatial resolution in advance and inherits whatever assumptions the atlas encodes. Learned tokenizers would instead discover spatial groupings from the data, potentially uncovering new spatial aggregation patterns beyond what is known anatomically. While MEG-GPT follows the parcellation approach, data-driven spatial tokenization has not been tried to date.

\subsection{Cross-modal integration with EEG and fMRI}

Several MEG-relevant foundation models already span modalities: BrainOmni pairs MEG with EEG, Brain-OF adds fMRI, and Ferrante et al.~\cite{ferrante_towards_2026} align all three. This direction is motivated by MEG's data limitations, and larger EEG and fMRI archives may complement MEG representations by introducing synergistic perspectives into brain dynamics. While brain foundation models are mainly still modality-specific, future approaches will likely integrate modalities in a more holistic brain model.

For now, end-to-end multi-modal training has been shown to underperform the best uni-modal model evaluated on its own modality~\cite{wang_what_2020}. Additionally, as data availability is highly asymmetric across MEG, EEG and fMRI, the direction of any gain cannot be taken for granted. None of the existing multi-modal models has been ablated against its own uni-modal counterpart, leaving two questions open: whether adding EEG or fMRI improves performance on MEG tasks relative to a MEG-only model, and whether adding MEG improves performance on EEG or fMRI tasks relative to models without it. Matched-protocol ablations in both directions would turn cross-modal integration from a plausible route into a supported one.

\subsection{From engineering metrics to neuroscientific insight}

While the above engineering questions are relevant to building better brain foundation models, their proper usage for scientific insight is yet to be established. Their primary application so far has been classification and regression of subject metadata and task condition or mental content. While these results are an impressive feat in brain decoding, their accuracy only establishes the presence of task-relevant information in a model's representations. It is still unclear which aspects of the underlying brain dynamics are carried across. A central ambition behind brain foundation models is to model cognition, and the risk is that they instead capture the measurement statistics of a particular modality, sensor array and preprocessing pipeline. The difficulty of cross-modal integration may be a symptom of exactly this: if the model’s internal representations were modeling cognition, data modalities would likely contribute synergistic perspectives onto the same dynamics rather than adding further noise statistics.

A deeper test would ask whether the learned representations reveal structure in cortical dynamics that was not already apparent. Candidate evidence includes latent dimensions that align with cognitive constructs without having been trained on them, representational geometry that recovers known functional organization, and, most compellingly, structure the model exposes that had not been described beforehand and can then be verified with conventional analyses. None of this is captured by a leaderboard, and the interpretability tooling for this is largely unbuilt for brain foundation models.

MEG provides a promising avenue here, combining the spatial fidelity of source-localized signals with millisecond temporal resolution. Source space also opens a route that sensor space does not: if the spatial aggregation is learned rather than imposed through an atlas parcellation, the model is free to discover spatial arrangements that do not follow known anatomical boundaries, and those groupings become an object of study in their own right. The best methodology remains an open question, and scaling law analyses across compute, model and data size are yet to be conducted for MEG.

\subsection{Data sharing, consent, and governance}

MEG foundation models will require data-sharing practices that are broader than those traditionally used for task-specific neuroimaging studies, but this expansion raises important questions of consent, stewardship, and governance. As Hanley et al.~\cite{hanley_training_2026} argue, brain foundation models create a new normative context because neural data are body-derived, historically collected under tightly governed clinical or research protocols, and now increasingly subject to large-scale repurposing, cross-context aggregation, and open-ended downstream use. These concerns are especially acute for MEG: the available datasets are relatively scarce, expensive to acquire, geographically concentrated, and often governed by heterogeneous access regimes. A responsible roadmap for MEG foundation models should therefore include explicit consent-compatibility review, dataset-level governance documentation, transparent rules for redistribution and derived models, mechanisms for withdrawal where feasible, and tiered-access frameworks that balance scientific reuse with privacy, participant dignity, and equitable benefit sharing.

\section{Conclusion}

The promise of MEG foundation models lies not in matching EEG or fMRI in data volume, but in exploiting the high quality and exceptional temporal and spatial richness of MEG recordings. MEG captures fast cortical dynamics at millisecond resolution, with relatively high spatial precision and, in typical whole-head systems, hundreds of simultaneously recorded channels. This combination can yield particularly information-rich datasets; recent brain-to-text work, for example, has reported stronger decoding performance with MEG than with EEG~\cite{jayalath_brains_2025}. Thus, even if MEG corpora remain comparatively smaller, their signal quality, dimensionality and spatial precision may make them particularly valuable for foundation-model pretraining. The field has arrived later to the foundation-model paradigm than neighbouring modalities, but it opens distinctive opportunities spanning source-resolved dynamics, naturalistic cognition, clinical brain-state inference and transfer across subjects, tasks and recording systems.

The design space remains largely open. Choices about model initialization, representation space, tokenization and pretraining objective are not merely implementation details: they determine what information the model can exploit, how well it can transfer across systems, and how interpretable its learned representations may become. Progress will therefore depend not only on larger and more diverse MEG corpora, but also on rigorous benchmarking using open, held-out datasets, standardized evaluation protocols and matched baselines. Given the scarcity of MEG data, corpus construction, benchmark design and responsible data sharing will need to be developed together.

The deeper question, however, is what these models should ultimately contribute to neuroscience and clinical practice. High decoding accuracy shows that task-relevant information is present in a representation; it does not establish that the model has learned meaningful or previously unknown structure in cortical dynamics, nor that it will improve clinically relevant decisions. A more demanding test is whether foundation models reveal latent organization that generates new hypotheses and can be independently verified, while also supporting more robust and transferable prediction of clinically meaningful outcomes, such as diagnosis, prognosis or treatment response. MEG is particularly well positioned for both goals because its combination of temporal precision and source-resolved spatial information provides a direct bridge between learned representations, brain dynamics and individual-level inference.

MEG foundation models therefore hold substantial promise across scientific, decoding and clinical applications. But realizing this promise will depend not only on better models and larger datasets, but also on how those datasets are assembled and used. As emphasized recently for brain foundation models more broadly, questions of consent, privacy, diversity, bias and data governance should be incorporated by design into data curation and model training, rather than addressed as an afterthought~\cite{hanley_training_2026}. The infrastructure, benchmarks and modeling principles for MEG foundation models are still being built, creating an opportunity to embed these considerations from the outset and develop models that are not only powerful, but robust, broadly applicable and responsibly built.

\section*{Acknowledgments}

This project was undertaken in part thanks to funding from IVADO and the Canada First Research Excellence Fund. K.J. is supported by funding from the Canada Research Chairs (950-232368) program and a Discovery Grant from the Natural Sciences and Engineering Research Council of Canada (2021-03426). This research was also made possible in part by support from the New Frontiers in Research Fund (NFRF) program through the Abundant Intelligences initiative. OPJ is supported by funding from MRC (MR/X00757X/1), Royal Society (RG\textbackslash{}R1\textbackslash{}241267), and ARIA (SCNI-SE01-P004).

\bibliographystyle{IEEEtran}
\bibliography{references}

\end{document}